\documentstyle[fullpage,epsf,twocolumn]{article}

\twocolumn
\title{Locality of the Density Matrix in Metals, Semiconductors and
Insulators}
\author{Sohrab Ismail-Beigi and T.A.~Arias\\
Department of Physics\\
Massachusetts Institute of Technology\\
Cambridge, MA 02139}

\date{\ }

\begin{document}

\maketitle

\begin{abstract}
We present an analytical study of the spatial decay rate
$\gamma$ of the one-particle density matrix $\rho(\vec r,\vec
r\,')\sim\exp(-\gamma|\vec r-\vec r\,'|)$ for systems described by
single particle orbitals in periodic potentials in arbitrary
dimensions.  This decay reflects electronic locality in condensed
matter systems and is also crucial for $O(N)$ density functional
methods.  We find that $\gamma$ behaves contrary to the conventional
wisdom that generically $\gamma\propto\sqrt{\Delta}$ in insulators and
$\gamma\propto\sqrt{T}$ in metals, where $\Delta$ is the {\em direct}
band gap and $T$ the temperature.  Rather, in semiconductors
$\gamma\propto\Delta$, and in metals at low temperature $\gamma\propto
T$.
\end{abstract}
\vspace{0.3in}

Density functional theory (DFT)\cite{HKS} describes many-body systems
via a single-particle formalism and is the basis for modern,
large-scale calculations in solid-state systems\cite{RMP}.  The
one-particle density matrix $\hat{\rho}\equiv\sum_n|\psi_n\rangle f_n
\langle\psi_n|$, which describes the state of a single-particle
quantum system, is the key quantity needed for the computation of
physical observables: total system energies, atomic forces, and
phonons can all be computed directly from $\hat\rho$.

Remarkably, despite the de-localized nature of the single particle
states $|\psi_n\rangle$, which may extend across an entire solid, the
physics of the electronic states in a given region of a material is
affected only by the local environment.  Reflecting this, the force on
an atom depends mostly on the positions of its nearest neighbors.
This electronic localization is manifest in the
``nearsightedness''\cite{K} of $\hat\rho$: $\rho(\vec r,\vec
r\,')\equiv\langle\vec r\,|\hat{\rho}|\vec r\,'\rangle \sim
\exp(-\gamma|\vec r-\vec r\,'|)$ where $\gamma>0$.  This exponential
decay has been verified numerically\cite{GT,SD} and
analytically\cite{KO,KR,DC1,GlowT}.

The locality of $\rho$ not only is important for understanding the
nearsightedness of effects arising from electronic structure but also
has direct practical impact on DFT calculations.  Recently, methods
have been proposed \cite{Yang,MGC,LNV,Daw,DS,SWF,GC,ODMG,HG,OAS} that
use $\rho$ directly and exploit its locality. Computationally, these
methods scale as $O(N)$, where $N$ is the number of atoms in the
simulation cell.  However, their prefactors depend strongly on
$\gamma$: some scale as $N/\gamma^6$\cite{LNV,Daw,HG} and others as
$N/\gamma^3$\cite{GC}.  Knowing how $\gamma$ depends on the system
under study is thus critical for carrying out such calculations.  For
a review of $O(N)$ methods, see \cite{GedeckerRMP}.

Generally, solid-state systems have an underlying periodic structure.
The introduction of localized defects\cite{KO} or surfaces\cite{KR}
does not change the spatial range of $\rho$ from that of the
underlying periodic lattice.  Thus, understanding the locality of
$\rho$ even for perfectly periodic systems is of direct relevance for
realistic material studies.  To date, the generic behavior of $\gamma$
is poorly understood.  For insulators in one dimension, Kohn has shown
that $\gamma\propto\sqrt{-E_n}$ in the tight-binding limit where $E_n$
is an atomic ionization energy\cite{Kohn}.  Motivated by this, it has
been assumed\cite{Kohn2,GT,ODMG} and argued\cite{Baer} that
$\gamma\propto\sqrt\Delta$ in multiple dimensions and more general
conditions, where $\Delta$ is the band gap.  For metals, it has been
assumed\cite{GT} and argued\cite{Baer} that $\gamma\propto\sqrt{T}$,
where $T$ is the electronic temperature.  However, the results
in\cite{Baer}, which to date represent the only effort to determine
$\gamma$ generically, are based on the assumption that the inverse of
the overlap matrix of a set of Gaussian orbitals decays in a Gaussian
manner. On the contrary, the inverses of such overlap matrices decay
only exponentially, and thus the behavior of $\gamma$ warrants further
study.

Here, we show that the behavior of $\gamma$ is more complex than
previously assumed.  For insulators, $\gamma$ is determined by the
analytical behavior of the filled bands, which is determined by the
strength of the periodic potential which, in turn, is most strongly
reflected in the size of the direct band gaps. Indirect gaps, being
more accidentally related to the strength of the potential, have a
more haphazard relation to $\gamma$, which we do not consider here.

For insulating systems, we find that $\gamma$ has the following
asymptotic behavior as a function of the {\em direct} gap $\Delta$,
lattice constant $a$, and electron mass $m$,
\begin{eqnarray}
\gamma \sim \left\{
\begin{array}{lll}
a\Delta m/\hbar^2\ \ \ & \mbox{for $a^2\Delta\rightarrow 0$}\ \ \
&\mbox{(weak-binding)}\\ \mbox{???}  & \mbox{for
$a^2\Delta\rightarrow\infty$} &\mbox{(tight-binding)}
\end{array}\right..\nonumber
\end{eqnarray}

The indeterminacy in the tight-binding limit results from the
electronic states becoming atomic orbitals, and thus, $\gamma$ depends
on the details of the underlying atomic potential.  Some systems (see
below) exhibit $\gamma\propto\sqrt{\Delta}$, but this is {\em not}
universal, as previously assumed.

\begin{figure}
\epsfxsize=3.2in
\epsfbox{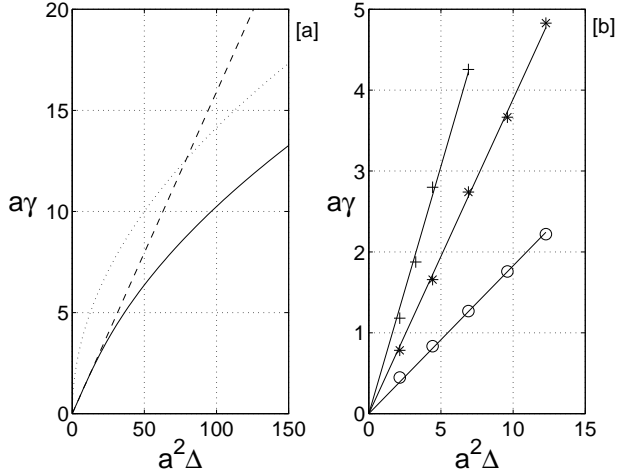}
\caption{[a] $a\gamma$ versus $a^2\Delta $ for a periodic array of
attractive delta potentials (solid curve).  The dashed line is
$a\gamma=a^2\Delta/(2\pi)$ . The dotted curve is
$a\gamma=a\sqrt{2\Delta }$, the leading asymptotic behavior of
$\gamma$ for large $a^2\Delta$. [b] $a\gamma$ versus $a^2\Delta$ for a
cubic lattice of Gaussian potentials: $\gamma$ in the [100] (circles),
[110] (stars) and [111] (pluses) directions.}
\label{gammacombined}
\end{figure}

One can, however, make a definitive statement in the weak-binding
limit, which is of direct importance for small-gap systems such as
those with weak pseudopotentials or gaps due to Jahn-Teller
distortions.  For example, semiconductors such as Si and GaAs have
gaps that are significantly smaller than their band widths, and we
expect them to fall into the weak-binding case.  In Si and GaAs, we
find $a^2\Delta m/\hbar^2\sim 4$ and $\sim 2.5$, respectively.
Inspecting Figure \ref{gammacombined}, we see that for such values the
behavior of $\gamma$ is well in the weakly-bound limit.

For metals with a fixed number of electrons, we find 
\begin{eqnarray}
\gamma \sim \left\{
\begin{array}{lll}
k_BT|\vec\nabla_k\,\varepsilon|^{-1}\ & \mbox{for}\ T\rightarrow 0\ 
& \mbox{(quantum)}\\
\left[{mk_B \over \hbar^2}T\ln\left({k_BT \over \varepsilon_F}\right)\right]^{1\over 2} &
\mbox{for}\ T\rightarrow\infty & \mbox{(classical)}
\end{array}\right.,\nonumber
\end{eqnarray}
in terms of the temperature $T$, the typical gradient of the band
energy $\varepsilon_{\vec k}$ on the Fermi surface, and the Fermi
energy $\varepsilon_F$.

The low-temperature result is of direct practical interest for
calculations in metals. (This result was also found in a
contemporaneously submitted publication \cite{GlowT}.)  We find
behavior resembling that proposed in \cite{GT,Baer},
i.e. $\gamma\propto\sqrt{T}$, only at extremely high temperatures.

We now present analytical arguments that substantiate the above
results and shed light on the physical mechanisms leading to and
differentiating among the different limits.  We consider periodic
systems with lattice vectors of characteristic length $a$.  We choose
units such that $\hbar^2/m=1$ and $k_B=1$ in order to avoid cumbersome
mathematical expressions; the results presented above are easily
recovered by inserting $\hbar^2/m$ and $k_B$, as appropriate, in each
step of the analysis below.

The Bloch wave-functions $\psi_{n\vec k}$, Wannier functions $W_n$,
Fermi-Dirac fillings $f_{n\vec k}$, and density matrix $\rho(\vec
r,\vec r\,')$ are related via
\begin{eqnarray}
W_n(\vec r,\vec R\,) & = & \Omega_B^{-1}\int d\vec k \,e^{-i\vec
k\cdot\vec R}\psi_{n\vec k}(\vec r\,)\nonumber \\
F_n(\vec R) & = & \Omega_B^{-1}\int d\vec k\, e^{i\vec
k\cdot\vec R} f_{n\vec k}\nonumber \\
\rho_n(\vec r,\vec r\,') & = & \sum_{\vec R} \sum_{\vec R'} W_n(\vec
r,\vec R\,)\,F_n(\vec R-\vec R')\,W_n^*(\vec r',\vec R')\nonumber \\
\rho(\vec r,\vec r\,') & = & \sum_n \rho_n(\vec r,\vec r\,').
\label{initdefs}
\end{eqnarray}
The integrals are over the first Brillouin zone with volume
$\Omega_B$. $\vec R$ ranges over the lattice vectors. $\rho_n(\vec
r,\vec r\,')$ is the density matrix of the $n$th band, and $\rho$ is a
simple sum over all $\rho_n$. Thus, we need only study the behavior of
$\rho_n$ for a given $n$.  We analyze the behavior of $\gamma$ for the
two possible cases of practical interest, insulators at low
temperature and metals at non-zero temperature.

{\em Insulators ($T=0$)} --- When the chemical potential $\mu$ falls
in the energy gap, all fillings $f_{\vec k}$ are 1 or 0.  A filled
band with $f_{\vec k}=1$ has $F(\vec R)=\delta_{\vec R,0}$ so that
$\rho$ is simply
\begin{eqnarray}
\rho(\vec r,\vec r\,') = \sum_{\vec R} W(\vec r,\vec R)W^*(\vec
r\,',\vec R).\nonumber
\end{eqnarray}
Wannier functions are exponentially
localized\cite{SD,Kohn,Kohn2,KO,KR,DC2,N,NN} and satisfy $W(\vec
r,\vec R\,)=W(\vec r-\vec R,0)$.  Thus, only a finite set of $\vec R$
contribute significantly to the above sum, and the decay rates of
$\rho$ and $W$ are the same.  Therefore, we need only determine
$\gamma$ for the Wannier functions.

As a concrete example and an initial orientation, we solve exactly for
$\gamma$ for the lowest band of a model one dimensional system for all
binding strengths.  We choose the periodic potential to be that of an
array of attractive delta-functions of strength $V>0$,
$U(x)=-V\sum_n\delta(x-na)$.  Following\cite{Kohn}, we define
$\mu(\varepsilon) \equiv \cos(a\sqrt{2\varepsilon
})-V\sin(a\sqrt{2\varepsilon})/\sqrt{2\varepsilon}$.  The band
structure is found by solving $\cos(ka) = \mu(\varepsilon_k)$ for real
$k$.  Focusing on the lowest band, we denote $\tilde\varepsilon$ as
the value of $\varepsilon$ where $\mu(\varepsilon)$ achieves its first
minimum. Kohn \cite{Kohn} has shown that
$\gamma=\cosh^{-1}|\mu(\tilde\varepsilon)|$.  We solve the above
transcendental system numerically for different values of $V$ and plot
$a\gamma$ as a function of $a^2\Delta$ in Figure \ref{gammacombined}[a].
The behavior at small $\Delta$ is clearly linear, showing that
$\gamma\sim a\Delta$ for a weak potential.  The leading asymptotic
behavior is $\gamma\sim\sqrt{\Delta}$ for a strong potential
(i.e. large $\Delta$).
In this case, the general notion that $\gamma\propto\sqrt\Delta$ is
clearly incorrect for weak potentials, and it is natural to ask
whether this result is peculiar to our simple model or whether it is
more universal.  As we now argue, for a weak potential, $\gamma\sim
a\Delta$ is quite general, whereas in the case of a strong potential,
the behavior of $\gamma$ is not unique and depends on the details of
the atomic system underlying the periodic lattice.  The crossover from
weak to strong potential behavior should occur when $\Delta$ is of
order of the band width.  In the figure, this occurs for $\Delta\sim
5(\pi/a)^2$.  We now analyze each case separately.

{\em Weak-binding insulators in general} --- We wish to find $\gamma$
in the limit of a weak periodic potential $U(\vec r\,)$.
Eqs. (\ref{initdefs}) show that the Wannier function is the Fourier
transform of $\psi_{\vec k}$.  Thus the range $\delta k$ in $\vec
k$-space where $\psi_{\vec k}$ has its strongest variations determines
the spatial range of $W$.  From basic Fourier analysis,
$\gamma\sim\delta k$.

A simple heuristic argument shows that $\gamma\sim a\Delta $.
Starting with a free electron description, a weak potential $U(\vec
r\,)$ causes the opening of a gap $\Delta$ at the edges of the
Brillouin zone.  The extent $\delta k$ of the region about the
zone-edges where $\varepsilon_{\vec k}$ deviates most appreciably from
its free electron value is given by $\delta k^2/2m^*\sim\Delta$, where
$m^*$ is the effective mass at the zone edge.  Standard
treatments\cite{AM} show that for weak potentials $m^*\sim a^2\Delta$.
Combining these results, we see that $\delta k\sim a\Delta$, whence
$\gamma\sim a\Delta$.

This heuristic argument gives the desired result, but there are hidden
assumptions.  The argument is based solely on the behavior of the band
structure $\varepsilon_{\vec k}$, whereas $W$ is determined by the
wavefunctions $\psi_{\vec k}$.  One must be sure that
$\varepsilon_{\vec k}$ and $\psi_{\vec k}$ vary over the same range
$\delta k$. Thus, We present a more precise argument in terms of
$\psi_{\vec k}$ alone.

Letting $\psi_{\vec k}(\vec r\,)=e^{i\vec k\cdot\vec r}u_{\vec k}(\vec
r\,)$, Eqs. (\ref{initdefs}) show that $W$ is also a Fourier transform
of $u_{\vec k}$.  Away from the edges of the Brillouin zone, $u_{\vec
k}$ is given perturbatively by
\begin{eqnarray}
u_{\vec k}(\vec r\,) = 1 + \sum_{\vec G\neq 0} {\langle\vec G|\hat
U|\vec 0\rangle\,e^{i\vec G\cdot\vec r} \over \left[k^2 - |\vec
k+\vec G|^2\right]/2} + O(U^2),\nonumber
\end{eqnarray}
where $\langle\vec r\,|\vec G\,\rangle=e^{i\vec G\cdot\vec r}$ and
$\vec G$ is a reciprocal lattice vector: $u_{\vec k}$ is smooth and
analytic in $\vec k$, and $u_{\vec k}\approx 1$.  However, close to
the zone edges, $|\vec k|\approx |\vec k+\vec G|$ and $u_{\vec k}$
deviates appreciably from unity in a region satisfying $[k^2-|\vec
k+\vec G|^2]/2\leq V$, where $V$ is the typical size of the matrix
elements of $U$. Since $|\vec G|\sim 1/a$, this region has a width
$\delta k\sim aV\sim a\Delta $ and hence $\gamma\sim a\Delta$.

As a concrete example, we study a cubic lattice of attractive Gaussian
potentials with rms width $a/\pi$.  We vary the depth of the
potential, and for each depth, we compute $\Delta$ and $\rho$ by
sampling the Brillouin zone on a cubic grid of size $40^3$ and
expanding $\psi_{\vec k}$ in plane waves with $|\vec G|\leq 12{\pi
\over a}$.  Diagonalizing the resulting Hamiltonian gives the
ground-state $\psi_{\vec k}$ from which we compute the density matrix.
Sampling $\rho(0,\vec r\,')$ in the $[100], [110]\ \mbox{and}\ [111]$
directions gives exponentially decaying envelopes upon which we
perform linear fits on log plots to extract $\gamma$. Figure
\ref{gammacombined}[b] shows our results, from which the behavior
$\gamma\propto\Delta$ is evident.

{\em Tight-binding insulators in general} --- The potential $U$ is the
periodic sum of an atomic potential $V_{at}$, $U(\vec r\,)=\sum_{\vec
R}V_{at}(\vec r-\vec R\,)$.  For sufficiently strong $V_{at}$, system
properties are determined by the atomic potential. The Wannier
functions become atomic orbitals localized about the minima of
$V_{at}$.  Now, $\gamma$ depends on the details of $V_{at}$ and no
single universal scaling can be found.  To demonstrate the complexity
and richness of this limit, we discuss briefly different examples of
atomic potentials that lead to differing forms for $\gamma$.  Note
that in this atomic limit the lattice constant $a$ is irrelevant in
determining $\gamma$.

For the Coulomb potential, $V_{at}(\vec r\,)=-Ze^2/r$. In the limit
$Ze^2\rightarrow\infty$, we have hydrogenic states centered on the
lattice sites with energies $E_n=-Z^2e^4/2n^2$ and Bohr radii
$a_0=n^2/Ze^2$.  The gap $\Delta$ is an energy difference between
atomic states, and so $\Delta\sim Z^2e^4$.  Also, $\gamma\sim
a_0^{-1}$, and so we conclude $\gamma\sim\sqrt{\Delta}$.  More
generally, for any atomic potential with only a single dimensionful
parameter (e.g. $Ze^2$ above), dimensional analysis gives
$\gamma\sim\sqrt{\Delta}$.

However, a similar analysis applied to a Gaussian potential,
$V_{at}(\vec r\,)=-Ve^{-r^2/2\sigma^2}$, gives
$\gamma\sim\Delta\sigma$, whereas, for a spherical well, $V_{at}(\vec
r\,)=-V\theta(\sigma-r)$, we find that $\gamma$ has {\em no}
dependence on $\Delta$.  Thus in the tight-binding limit, it is
difficult to make generic statements regarding $\gamma$.

{\em Metals ($T>0$)} --- In metals, the fillings $f_{\vec k}$ exhibit
rapid variations in $\vec k$ across the Fermi surface and $F(\vec
R\,)$ in (\ref{initdefs}) becomes long-ranged.  The Wannier functions,
being independent of the fillings, remain exponentially localized, as
discussed above.  These facts combined with the structure of the sum
in (\ref{initdefs}) imply that $\gamma$ in this case is determined by
$F(\vec R\,)$.
\begin{figure}
\epsfxsize=2.8in
\epsfbox{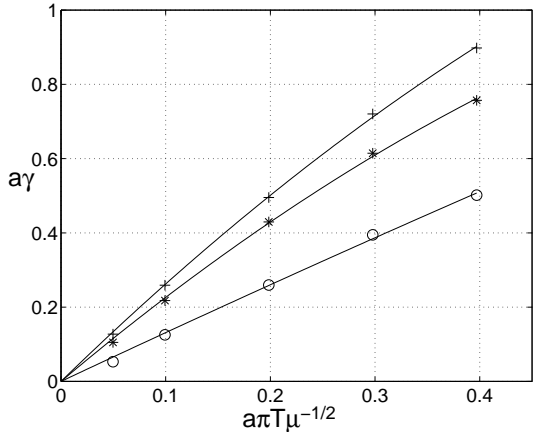}
\caption{$a\gamma$ versus $a\pi T\mu^{-{1\over 2}}$ for a BCC lattice
of tight-binding s-orbitals: $\gamma$ in the [100] (circles), [110]
(stars) and [111] (pluses) directions. $\mu$ is measured from the
bottom of the band (see text).}
\label{gammabcc}
\end{figure}
For an initial orientation, consider a band with a free electron-like
form $\varepsilon_k=k^2/2$, whose spherical Fermi surface is contained
inside the first Brillouin zone.  For such a band, $F(\vec R\,)$ is
given by (\ref{initdefs}) with
$f_k=1/\left(1+\exp\left[(k^2/2-\mu)/T\right]\right)$.  Below, we will
use the fact that the density matrix of a true free electron gas is
proportional to this $F$: $\rho(\vec r,\vec r\,')\propto F(\vec r-\vec
r\,')$.

Because the Fermi surface is contained within the first zone, we may
extend the $\vec k$ integral for $F$ to infinity.  Changing to
spherical coordinates, integrating by parts and using trigonometric
identities yields
\begin{equation}
F(\vec R\,) =  {1\over 2\Omega_BT}\left({1 \over R}{\partial \over
\partial R}\right)^2\int_{-\infty}^\infty{dk\,\cos\left(kR\right)
\over \cosh^2\left[{k^2/2-\mu \over 2T}\right]}.
\label{freeelec}
\end{equation}
When closing the integral in the upper complex $k$-plane, the relevant
poles of the integrand are at $k=\tilde k_l\equiv\pm\sqrt{2\mu\pm
2i\pi T(2l+1)}$ for integers $l\ge 0$.  The residues of these poles
contain the factor $e^{i\tilde k_lR}$ which gives rise to oscillations
due to the real part of $\tilde k_l$ and exponential decay due to its
imaginary part.

When $T\rightarrow 0$, $\mu$ equals the Fermi energy
$\mu=\varepsilon_F=k_F^2/2$ where $k_F=(3\pi^2n)^{1/3}$ and $n$ is the
electron density.  Thus we have $\tilde k_l\approx i\pi T(2l+1)/k_F\pm
k_F$.  As $R\rightarrow\infty$, the $l=0$ contribution dominates, so
that $\gamma=\pi T/k_F$.

As $T\rightarrow\infty$, the ideal gas result $\mu\approx
T\ln(n\lambda_T^3)$ holds where $\lambda_T=\sqrt{2\pi/T}$ is the
thermal de Broglie wavelength.  In this limit, $\tilde
k_l\approx\sqrt{2\mu}$ so that
$\gamma=Im\sqrt{2T\ln(n\lambda_T^3)}\sim\sqrt{T\ln(T/\varepsilon_F)}$.

Note that if we approximate the integrand of Eq. (\ref{freeelec}) by
$e^{\mu/T-k^2/2T}$, which corresponds to using Maxwell-Boltzmann
fillings, $F$ will have a Gaussian form $F(\vec R\,)\propto
e^{-TR^2/2}$.  However, one can show that this approximation is only
valid for small $R$.  For large $R$, an exponential tail $e^{-\gamma
R}$ remains where $\gamma$ is as described above.

We now present separate arguments showing that these asymptotic forms
for $\gamma$ are correct for metals in general.

{\em Metals as $T\rightarrow 0$} --- We first consider first $T=0$.
Bands below the Fermi level then have $f_{\vec k}=1$ and $F(\vec
R\,)=\delta_{\vec R,0}$, and, as for the insulating case, their
density matrices decay exponentially.  However, for bands that cross
the Fermi level, $f_{\vec k}$ jumps discontinuously from unity to zero
wherever $\varepsilon_{\vec k}=\mu$.  As is well known, the Fourier
transform of a discontinuous function has algebraic falloff, and thus
$F(\vec R\,)|_{T=0}\propto |\vec R\,|^{-\eta}$ where $\eta>0$.  Such
bands therefore dominate the decay of $\rho$ as $T\rightarrow 0$.

At finite $T$, the fillings are $f_{\vec k}=\left(1+e^y\right)^{-1}$
where $y=\left(\varepsilon_{\vec k}-\mu\right)/T$.  As $T\rightarrow
0$, the fillings now go from unity to zero in a narrow region about
the Fermi surface defined by vectors $\vec k_F$ satisfying
$|\varepsilon_{\vec k}-\mu|\sim T$. To determine the width of this
region, we approximate $f_{\vec k}$ about the Fermi vector $\vec k_F$
via $y\approx\vec\nabla\varepsilon\cdot(\vec k-\vec k_F)/T$.  The
width of the transition region and $\gamma$ thus are given by
$\gamma\sim\delta k\sim T/|\vec\nabla\varepsilon|$.  This argument
holds for any Fermi surface no matter how complex (metallic or
semi-metallic) at sufficiently low $T$ such that $\delta k$ is smaller
than the typical scale of the features of the Fermi surface.

As a further verification for the general case, we study a
body-centered cubic lattice of tight-binding s-orbitals with lattice
constant $a=4.32$\AA, for which the Fermi surface is non-spherical.
We choose the tight-binding matrix element so that the band structure
has the free electron effective mass at $\vec k=0$.  Choosing $\mu$ to
be $2/5$ of the way from the band minimum to the band maximum, we
calculate $F(\vec R\,)$ for various values of $T$.  Sampling the
Brillouin zone on a $200^3$ grid, plotting $F(\vec R\,)$ and finding
exponentially decaying envelopes, we perform linear fits on a log plot
and extract $\gamma$. Figure \ref{gammabcc} shows that indeed
$\gamma\propto T$ for $T\rightarrow 0$.

{\em Metals as $T\rightarrow\infty$} --- Here the electron kinetic
energy is much larger than the periodic potential so that we may
approximate $\varepsilon_{\vec k}=k^2/2$: the system is a classical
ideal gas and is not of interest for solid-state calculations.  Our
previously derived result for a free electron gas yields
$\gamma\sim\sqrt{T\ln(T/\varepsilon_F)}$.  Only in this limit do we
find a result resembling that of \cite{Baer}.

We thank S. Goedecker and D. Vanderbilt for discussions
and helpful criticisms.  This work was supported primarily by the
MRSEC Program of the National Science Foundation under award number
DMR 94-00334 and also by the Alfred P. Sloan Foundation (BR-3456).

\end{document}